\documentclass[onecolumn]{IEEEtran}
\IEEEoverridecommandlockouts
\usepackage{cite}
\usepackage{amsmath,amssymb,amsfonts}
\usepackage{algorithmic}
\usepackage{graphicx}
\usepackage{textcomp}
\usepackage{xcolor}
\usepackage{graphicx}
\usepackage{subcaption}
\usepackage{tabularx}
\usepackage{float} 
\usepackage{makecell}
\newcommand{\thickhline}{\Xhline{2\arrayrulewidth}}
\usepackage{multirow}
\usepackage{color}
\usepackage{arydshln} 
\usepackage[utf8]{inputenc}
\usepackage{tcolorbox}
\definecolor{Color1}{RGB}{128, 0, 128}   
\definecolor{Color2}{RGB}{150, 150, 0}    
\definecolor{Color3}{RGB}{0, 180, 180}    
\definecolor{Color4}{RGB}{100,50,150} 
\usepackage{booktabs} 
\usepackage{algorithmic}
\usepackage{blindtext}
\usepackage{titlesec}

\def\BibTeX{{\rm B\kern-.05em{\sc i\kern-.025em b}\kern-.08em
    T\kern-.1667em\lower.7ex\hbox{E}\kern-.125emX}}
    
\begin{document}

\title{In-Depth Analysis of Emotion Recognition through Knowledge-Based Large Language Models}


\author{%
    \begin{tabular}[t]{@{}c@{}} 
        \IEEEauthorblockN{Bin Han}\\
        \IEEEauthorblockA{ USC ICT\\
        Los Angeles, CA, USA\\
        binhan@usc.edu}
    \end{tabular}
    \hspace*{5pt} 
    \begin{tabular}[t]{@{}c@{}}
        \IEEEauthorblockN{Cleo Yau}\\
        \IEEEauthorblockA{ California State Polytechnic\\
        Pomona, CA, USA\\
        cleoyau@gmail.com}
    \end{tabular}
    \hspace*{5pt}
    \begin{tabular}[t]{@{}c@{}}
        \IEEEauthorblockN{Su Lei}\\
        \IEEEauthorblockA{audEERING GmbH\\
        Gilching, Germany\\
        slei@audeering.com}
    \end{tabular}
    \hspace*{5pt}
    \begin{tabular}[t]{@{}c@{}}
        \IEEEauthorblockN{Jonathan Gratch}\\
        \IEEEauthorblockA{ USC ICT\\
        Los Angeles, CA, USA\\
        gratch@ict.usc.edu}
    \end{tabular}
}

\maketitle

\tableofcontents

\vspace{3em}

\begin{abstract}
Emotion recognition in social situations is a complex task that requires integrating information from both facial expressions and the situational context. 
While traditional approaches to automatic emotion recognition have focused on decontextualized signals, recent research emphasizes the importance of context in shaping emotion perceptions. 
This paper contributes to the emerging field of context-based emotion recognition by leveraging psychological theories of human emotion perception to inform the design of automated methods. 
We propose an approach that combines emotion recognition methods with Bayesian Cue Integration (BCI) to integrate emotion inferences from decontextualized facial expressions and contextual knowledge inferred via Large-language Models. 
We test this approach in the context of interpreting facial expressions during a social task, the prisoner's dilemma. 
Our results provide clear support for BCI across a range of automatic emotion recognition methods. The best automated method achieved results comparable to human observers, suggesting the potential for this approach to advance the field of affective computing. 

\end{abstract}

\vspace{2em}

\section{Overview}
In this work, we propose a general approach to context-dependent emotion recognition~\cite{han2024knowledge}.
This approach consists of three main steps.
The first step is to predict the emotions that people are likely to perceive from an emotional expression without context (see Section~\ref{sec:Pef}).
The second step involves predicting the emotions that people are likely to perceive from a situational description.
Finally, we combine these separate sources of information using psychologically-inspired models, such as Bayesian Cue Integration~\cite{ong2015affective} (see Section~\ref{sec:pecf}).
We test this approach with an emotional social task called the prisoner's dilemma (game details and corpus are explained in~\cite{han2024knowledge}).

\newpage

\section{Emotion Probability given by Face}
\label{sec:Pef}
This section focuses on emotion distribution estimated from facial cues alone. We compare three alternatives for automatically recognizing emotions from decontextualized videos. 

\subsection{LSTM Model Details}
We train an LSTM model that incorporates dynamic features using human context-free annotations as the ground truth. The model utilizes various input features, including sequences of Action Units (AUs), facial optical flows, gaze, and head pose data. Detailed descriptions of the model architecture are provided in Table~\ref{table:lstm_detailed}.

\vspace{1.5em}

\label{LSTM_Architecture}
\begin{table}[h!]
  \centering
    \setlength{\tabcolsep}{15pt} 
  \resizebox{0.5\textwidth}{!}{
    \begin{tabular}{cc}
      \toprule
      Layer Type & Configuration \\
      \midrule
      LSTM & return sequences=True, units=50 \\
      Dropout & rate=0.5 \\
      LSTM & units=50 \\
      Dropout & rate=0.5 \\
      Dense & units=7, activation=`linear' \\
      \bottomrule
    \end{tabular}
    
  }
  \caption{LSTM Model Architecture}
   \label{table:lstm_detailed}
\end{table}

\subsection{Emotional Probability Distribution: $P(e|f)$}

We treat context-free human annotations as ground truth (human annotation results) when people judge emotions without situational context.
Figure~\ref{fig:three_graphs} shows the emotional probability distribution for three facial emotion recognition methods: FACET, EAC, and LSTM. 
FACET and EAC overestimate joy across all conditions. FACET and EAC do not recognize surprise well, which could significantly affect their performance (see Table 3 in~\cite{han2024knowledge}).

\begin{figure*}[h]
    \centering
    \begin{subfigure}[b]{0.32\textwidth}
        \centering
        \includegraphics[width=\textwidth]{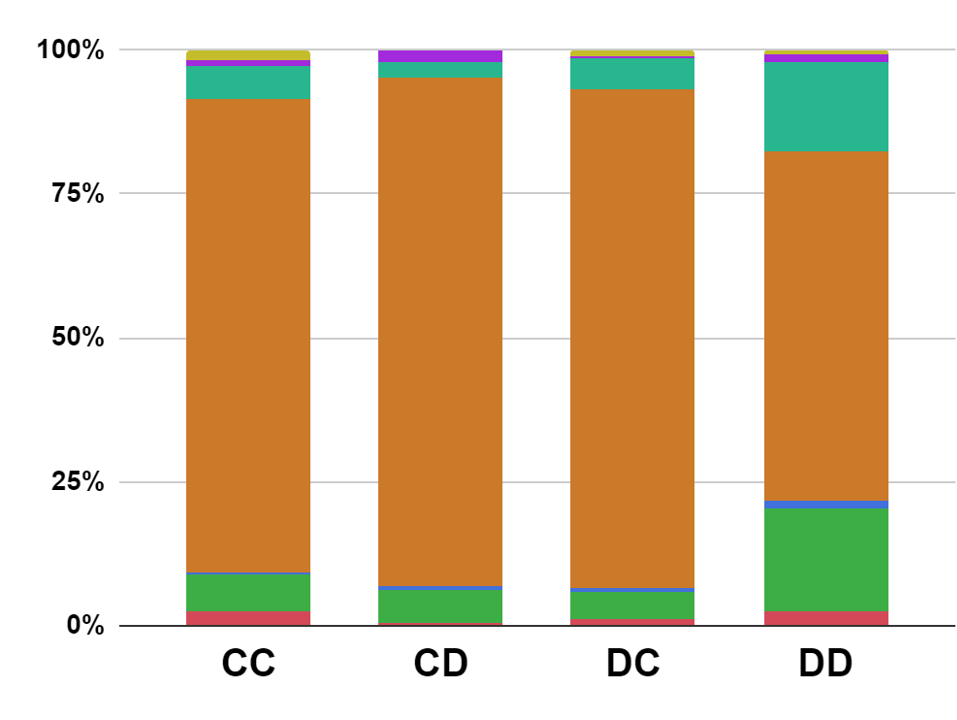}
        \caption{FACET}
        \label{fig:zerocontext}
    \end{subfigure}
    \begin{subfigure}[b]{0.32\textwidth}
        \centering
        \includegraphics[width=\textwidth]{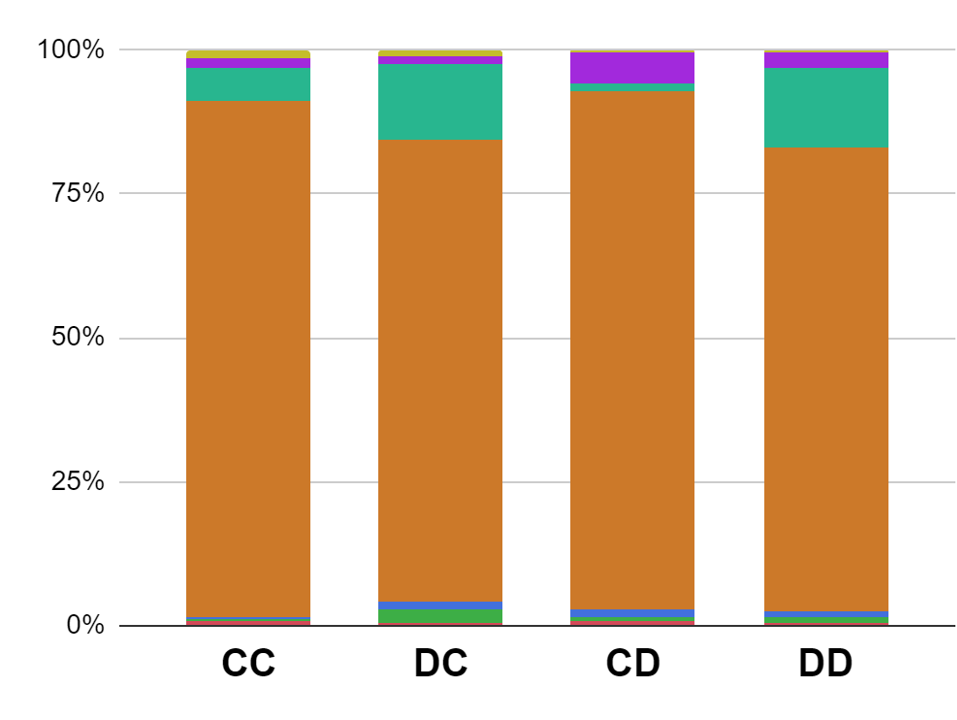}
        \caption{EAC}
        \label{fig:withcont}
    \end{subfigure}
    \begin{subfigure}[b]{0.32\textwidth}
        \centering
        \includegraphics[width=\textwidth]{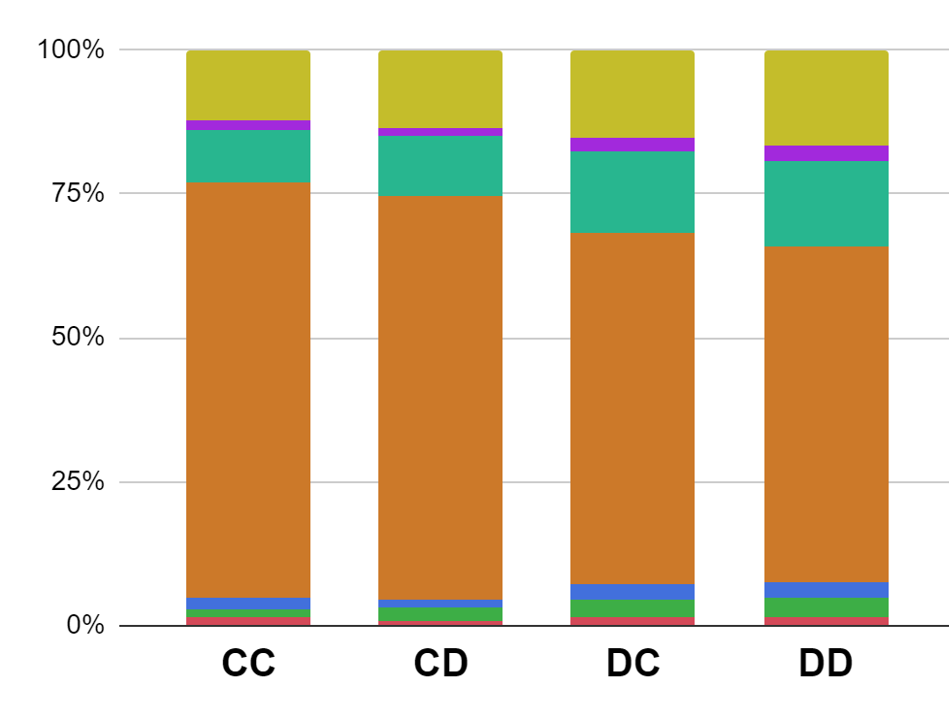}
        \caption{LSTM}
        \label{fig:human_novideo}
    \end{subfigure}
    \caption{Emotional probability distribution for $P(e|f)$.}
\label{fig:three_graphs}
\end{figure*}


\section{Emotion Probability given by Face and Context}
\label{sec:pecf}
We compare alternative methods for integrating facial and contextual cues for context-aware emotion recognition. We explore three methods: BCI, GPT-4, and NN.

\vspace{2em}

\subsection{GPT Prompt for Integration}

\begin{figure}[h]
    \centering
    \includegraphics[width=1\linewidth]{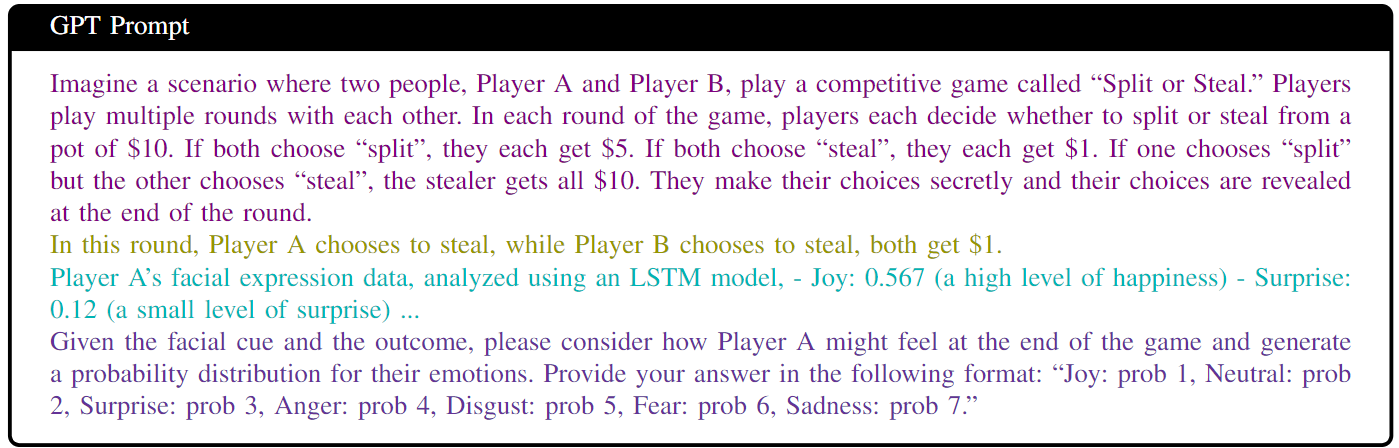}
    \caption{GPT Prompt for $P(e|c,f)$: Integrating facial cue and the game outcome to predict Player A's emotion.}
    \label{fig:enter-label}
\end{figure}

LLMs are known for their ability to understand various situational cues and perform well, especially in terms of emotional capability~\cite{tak2023gpt}. We provided the LLM with situational knowledge and outcome to obtain emotional probabilities. 
This method utilizes the advanced capabilities of GPT-4 to directly generate a context-aware emotion probability distribution. 
The integration process involves GPT-4, which incorporates a representation of $P(e|f)$ estimated by a context-free facial emotion recognition method.
Additionally, we crafted prompts to integrate both context and face cues to the LLM.
We used GPT-4.~\footnote{GPT-4 versions as of February 1, 2024}.
The Figure~\ref{fig:enter-label} shows the GPT prompt. The prompt contains four main components:
\begin{itemize}
    \item \textcolor{Color1}{A general description of the prisoner's dilemma game~\cite{rapoport1965prisoner}.}
    \item \textcolor{Color2}{The game outcome of each turn (CC,DC,CD, and DD).}
    \item \textcolor{Color3}{Representation of $P(e|f)$ estimated by a LSTM.}
    \item \textcolor{Color4}{A request for the emotional distribution (Basic emotion).}
\end{itemize}

\subsection{Neural Network Integration}

In this section, we discuss the Neural Network Integration (NNI) method. To improve integration, a simple neural network was trained using LSTM emotion distribution results ($P(e|f)$) and GPT-4 emotion distribution results ($P(e|c)$) as inputs, with context-based results ($P(e|c,f)$) as the ground truth. 
The NN architecture included a dense layer with 100 ReLU neurons and a softmax output layer. 
Using the Adam optimizer and a custom KL Divergence loss function, the model was trained over 1000 epochs. Cross-validation with 5 splits ensured robust evaluation. 
Table~\ref{tab:comparison_results} shows the result of the comparison. 
To assess the performance of models, we employ three standard metrics. 
KLD~\cite{kullback1951information} and RMSE are standard metrics to compare the distance between two probability distributions~\cite{peng2015mixed, zhao2017approximating}.
Additionally, we adopt F1 (weighted) to evaluate
performance if the model was forced to provide a single label.
The NN's performance was somewhat worse than the BCI and GPT-4 integration results. It was only compared with the LSTM because both were fine-tuned on split-steal data, unlike the pre-trained models.

\vspace{2em}

\begin{table}[h]
    \centering
     \setlength{\tabcolsep}{18pt}  
    \resizebox{0.8\textwidth}{!}{%
    \begin{tabular}{l||ccc}
     \thickhline
    \textbf{Face+Context (Integration)} & \textbf{KLD(↓)} & \textbf{RMSE(↓)} & \textbf{F1(↑)} \\
    \hline
    FACET+GPT-3 (BCI) & 1.713 & 0.215 & 0.525 \\
    FACET+GPT-4 (BCI) & 1.829 & 0.200& 0.565 \\
    EAC+GPT-3 (BCI) & 1.340 & 0.200 & 0.519\\
    EAC+GPT-4 (BCI) & 1.330 & 0.210 &0.527\\
    LSTM+GPT-3 (BCI) & 0.809 & 0.162& 0.454\\
    LSTM+GPT-4 (BCI) & \textbf{0.346} & 0.104&0.649 \\
    Human+Human (BCI) & 0.441 & \textbf{0.092}& \textbf{0.782} \\
    \hdashline
    FACET (GPT-4) & 0.648  & 0.150&0.528 \\
    EAC (GPT-4) & 0.597 & 0.155&0.503 \\
    LSTM (GPT-4) & 0.354 & 0.112&0.530 \\
    \hdashline
    LSTM+GPT-4 (NNI) &  0.580  &  0.151 &  0.151 \\
    \hline
    \end{tabular}
    }
    \caption{Comparison of alternative integration methods for $P(e|c,f)$ using different combinations of facial (FACET, EAC, LSTM) and context (GPT-3 and GPT-4) and integration methods (BCI and GPT-4). The best results for each measure}
    \label{tab:comparison_results}
\end{table}

\subsection{Emotional Probability Distribution: $P(e|c,f)$}

Fig.~\ref{fig:three_graphs} illustrates the emotional distributions when integrating facial and contextual cue for emotion recognition. Methods integrating FACET and EAC show a propensity to overestimate joy in the CD. 
In addition, integrations with GPT-4 yield distributions that are more closely aligned with human context-based estimations than those with GPT-3.

\begin{figure*}[h]
    \centering
    \begin{subfigure}[b]{0.32\textwidth}
        \centering
        \includegraphics[width=\textwidth]{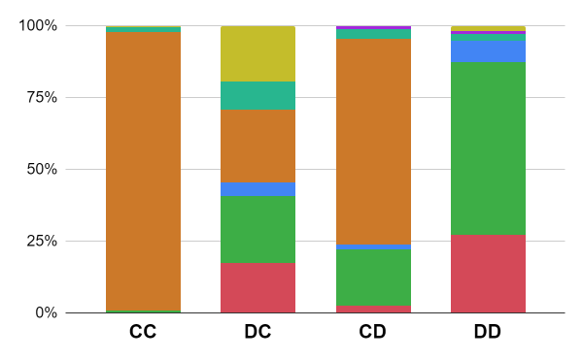}
        \caption{FACET+GPT-3(BCI)}
        \label{fig:zerocontext}
    \end{subfigure}
    \begin{subfigure}[b]{0.32\textwidth}
        \centering
        \includegraphics[width=\textwidth]{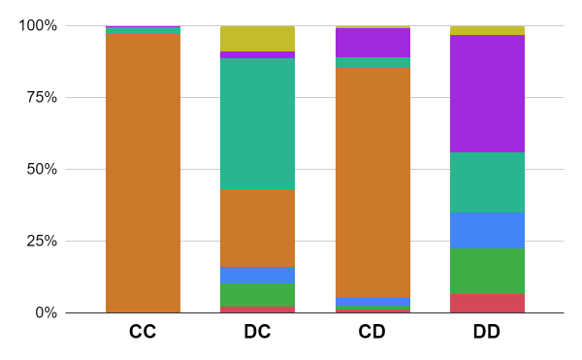}
        \caption{EAC+GPT-3(BCI)}
        \label{fig:withcont}
    \end{subfigure}
    \begin{subfigure}[b]{0.32\textwidth}
        \centering
        \includegraphics[width=\textwidth]{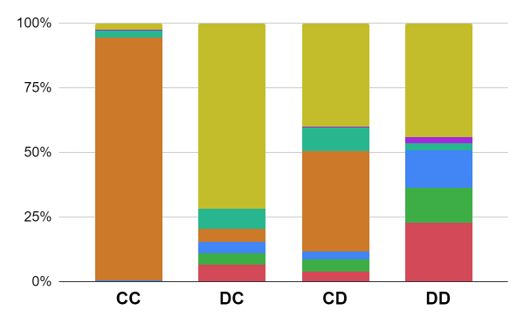}
        \caption{LSTM+GPT-3 (BCI)}
        \label{fig:human_novideo}
    \end{subfigure}
    \\
        \centering
    \begin{subfigure}[b]{0.32\textwidth}
        \centering
        \includegraphics[width=\textwidth]{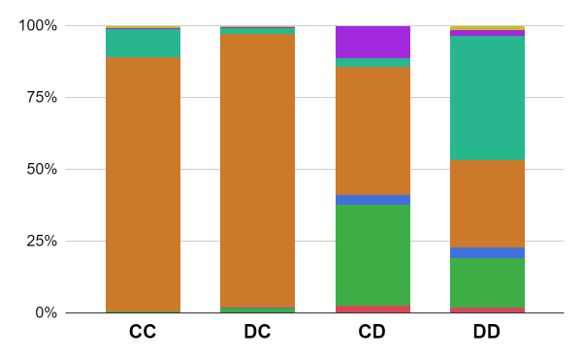}
        \caption{FACET+GPT-4 (BCI)}
        \label{fig:zerocontext}
    \end{subfigure}
    \begin{subfigure}[b]{0.32\textwidth}
        \centering
        \includegraphics[width=\textwidth]{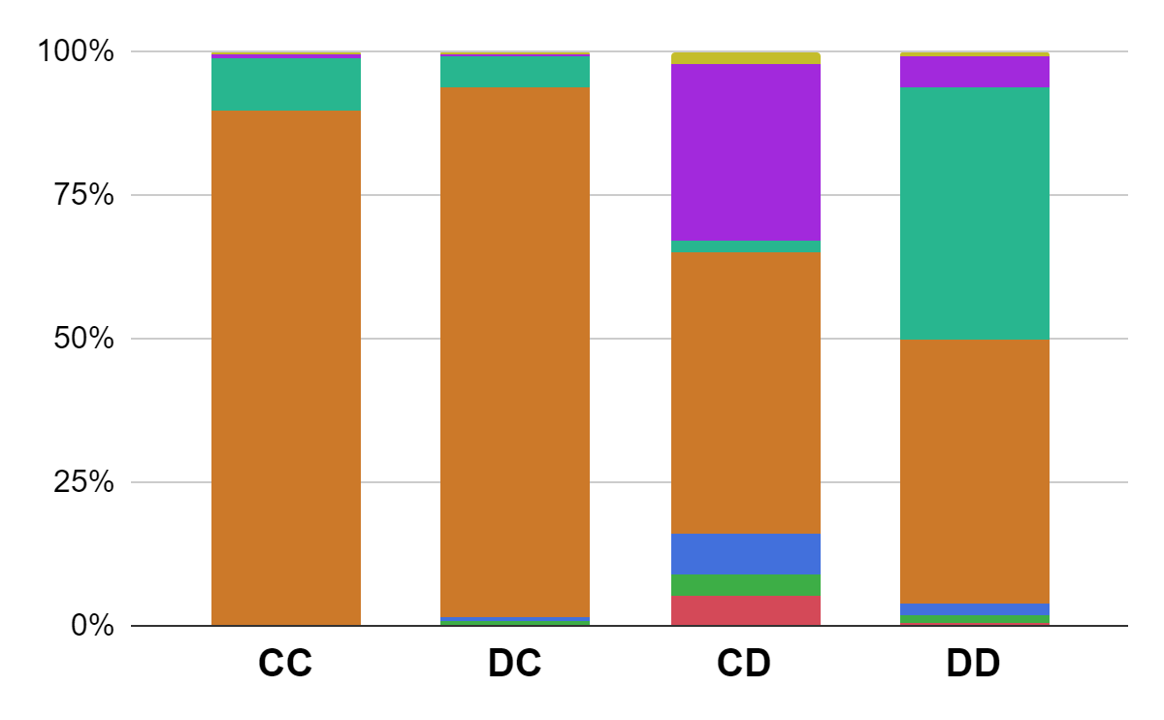}
        \caption{EAC+GPT-4 (BCI)}
        \label{fig:withcont}
    \end{subfigure}
    \begin{subfigure}[b]{0.32\textwidth}
        \centering
        \includegraphics[width=\textwidth]{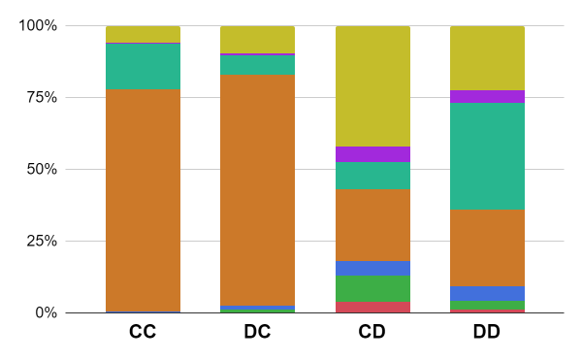}
        \caption{LSTM+GPT-4 (BCI)}
        \label{fig:human_novideo}
    \end{subfigure}
    \\
    \centering
    \begin{subfigure}[b]{0.32\textwidth}
        \centering
        \includegraphics[width=\textwidth]{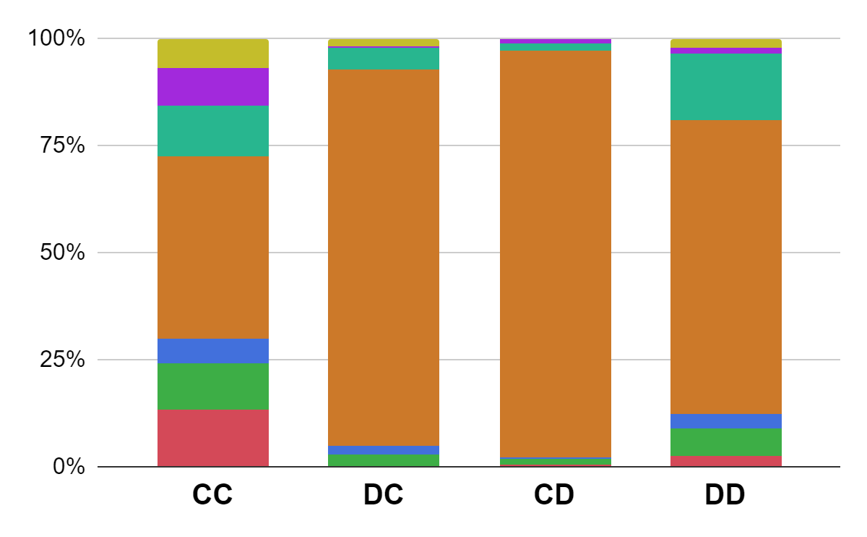}
        \caption{FACET(GPT-4)}
        \label{fig:zerocontext}
    \end{subfigure}
    \begin{subfigure}[b]{0.32\textwidth}
        \centering
        \includegraphics[width=\textwidth]{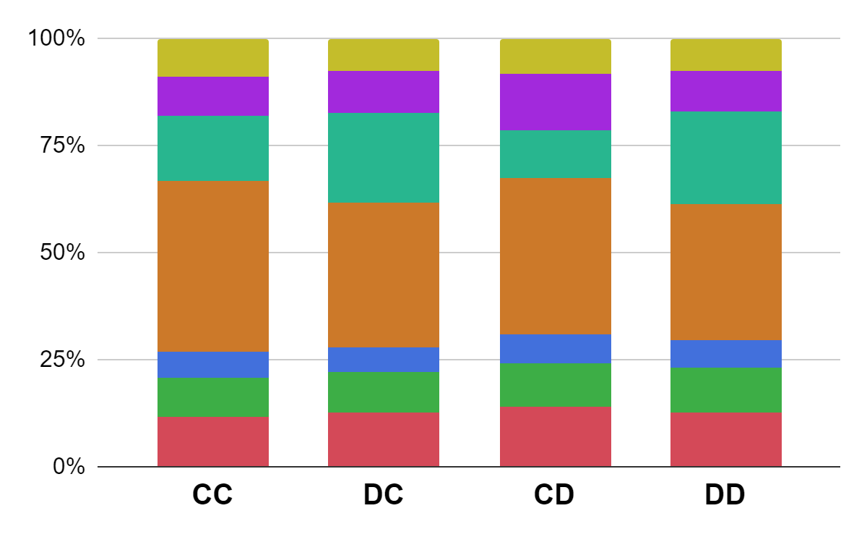}
        \caption{EAC(GPT-4)}
        \label{fig:withcont}
    \end{subfigure}
    \begin{subfigure}[b]{0.32\textwidth}
        \centering
        \includegraphics[width=\textwidth]{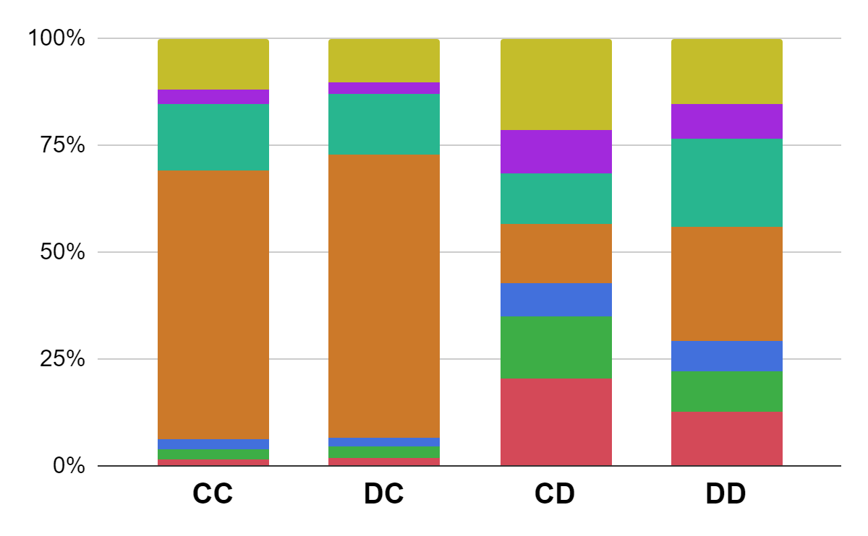}
        \caption{LSTM(GPT-4)}
        \label{fig:human_novideo}
    \end{subfigure}
    \caption{Emotional probability distribution for $P(e|c,f)$.}
    \label{fig:three_graphs}
\end{figure*}

\section{Performance}
\subsection{Evaluation by Context (KLD, RMSE, and F1-Score)}

We assess the performance of knowledge-based emotion recognition against human context-based perception distributions. 
The LSTM+GPT-4 (BCI) shows robust performance across all contexts, mirroring the trend observed in human-to-human (BCI) assessments. 
Notably, most methods displayed weaker performance in CD and DD contexts, whereas LSTM+GPT-4 (BCI) and human-to-human (BCI) demonstrated relatively better performance in these contexts.


\begin{table}[h]
\centering
\setlength{\tabcolsep}{10pt} 
\renewcommand{\arraystretch}{0.99} 
\resizebox{0.8\textwidth}{!}{%
\begin{tabular}{l||r|rrrr}
\thickhline
 & \textbf{Overall} & \textbf{CC} & \textbf{DC} & \textbf{CD} & \textbf{DD} \\
\hline
FACET+GPT-3 (BCI) & 1.713 & 1.301 & 1.158 & 2.407 & 1.986 \\
FACET+GPT-4 (BCI) & 1.829 & 1.147 & 1.946 & 1.948 & 2.275 \\
EAC+GPT-3 (BCI) & 1.341 & 1.152 & 0.819 & 2.024 & 1.367 \\
EAC+GPT-4 (BCI) & 1.330 & 0.886 & 1.486 & 1.495 & 1.453 \\
LSTM+GPT-3 (BCI) & 0.809 & 0.394 & 0.935 & 0.385 & 1.521 \\
LSTM+GPT-4 (BCI)& 0.346 & 0.222 & 0.376 & 0.389 & 0.395 \\
Human+Human (BCI) & 0.441 & 0.235 & 0.379 & 0.59 & 0.561 \\
\hdashline
FACET(GPT-4) & 0.648 & 0.549 & 0.557 & 0.663 & 0.824 \\
EAC(GPT-4) & 0.597 & 0.491 & 0.688 & 0.621 & 0.588 \\
LSTM(GPT-4) & 0.354 & 0.179 & 0.291 & 0.538 & 0.407 \\
\hline
\end{tabular}
}
\caption{Break down the KLD result by context}
\end{table}


\begin{table}[h]
\centering
\setlength{\tabcolsep}{10pt} 
\renewcommand{\arraystretch}{1.1} 
\resizebox{0.8\textwidth}{!}{%
\begin{tabular}{l||r|rrrr}
\thickhline
 & \textbf{Overall} & \textbf{CC} & \textbf{DC} & \textbf{CD} & \textbf{DD} \\
\hline
FACET+GPT-3 (BCI) & 0.215 & 0.134 & 0.189 & 0.272 & 0.267 \\
FACET+GPT-4 (BCI) & 0.2 & 0.132 & 0.175 & 0.254 & 0.238 \\
EAC+GPT-3 (BCI)& 0.2 & 0.137 & 0.163 & 0.282 & 0.218 \\
EAC+GPT-4 (BCI)& 0.21 & 0.138 & 0.194 & 0.26 & 0.251 \\
LSTM+GPT-3 (BCI)& 0.162 & 0.097 & 0.241 & 0.108 & 0.202 \\
LSTM+GPT-4 (BCI)& 0.104 & 0.068 & 0.113 & 0.116 & 0.117 \\
Human+Human (BCI) & 0.092 & 0.056 & 0.088 & 0.106 & 0.117 \\
\hdashline
FACET(GPT-4) & 0.15 & 0.15 & 0.14 & 0.153 & 0.159 \\
EAC(GPT-4) & 0.155 & 0.1484 & 0.1626 & 0.1518 & 0.1587 \\
LSTM(GPT-4) & 0.112 & 0.076 & 0.098 & 0.145 & 0.131 \\
\hline
\end{tabular}
}
\caption{Break down the RMSE result by context}
\end{table}


\begin{table}[h]
\centering
\setlength{\tabcolsep}{10pt} 
\renewcommand{\arraystretch}{1.1} 
\resizebox{0.8\textwidth}{!}{%
\begin{tabular}{l||r|rrrr}
\thickhline
 & \textbf{Overall} & \textbf{CC} & \textbf{DC} & \textbf{CD} & \textbf{DD} \\
\hline
FACET+GPT-3 (BCI) & 0.525 & 0.882 & 0.676 & 0.333 & 0.300 \\
FACET+GPT-4 (BCI) & 0.565 & 0.861 & 0.838 & 0.375 & 0.338 \\
EAC+GPT-3 (BCI) & 0.519 & 0.861 & 0.711 & 0.324 & 0.181 \\
EAC+GPT-4 (BCI) & 0.527 & 0.861 & 0.773 & 0.323 & 0.328 \\
LSTM+GPT-3 (BCI)& 0.454 & 0.882 & 0.045 & 0.390 & 0.058 \\
LSTM+GPT-4 (BCI)& 0.649 & 0.861 & 0.804 & 0.192 & 0.510 \\
Human+Human (BCI) & 0.782 & 0.954 & 0.795 & 0.753 & 0.602 \\
\hdashline
FACET(GPT-4) & 0.528 & 0.861 & 0.782 & 0.343 & 0.289 \\
EAC(GPT-4) & 0.503 & 0.861 & 0.750 & 0.333 & 0.267 \\
LSTM(GPT-4) & 0.530 & 0.861 & 0.773 & 0.142 & 0.333 \\
\hline
\end{tabular}

}
\caption{Break down the RMSE result by context}
\end{table}

\newpage
\subsection{Confusion Matrix}

\begin{figure*}[h]
    \centering
    \begin{subfigure}[b]{0.26\textwidth}
        \centering
        \includegraphics[width=\textwidth]{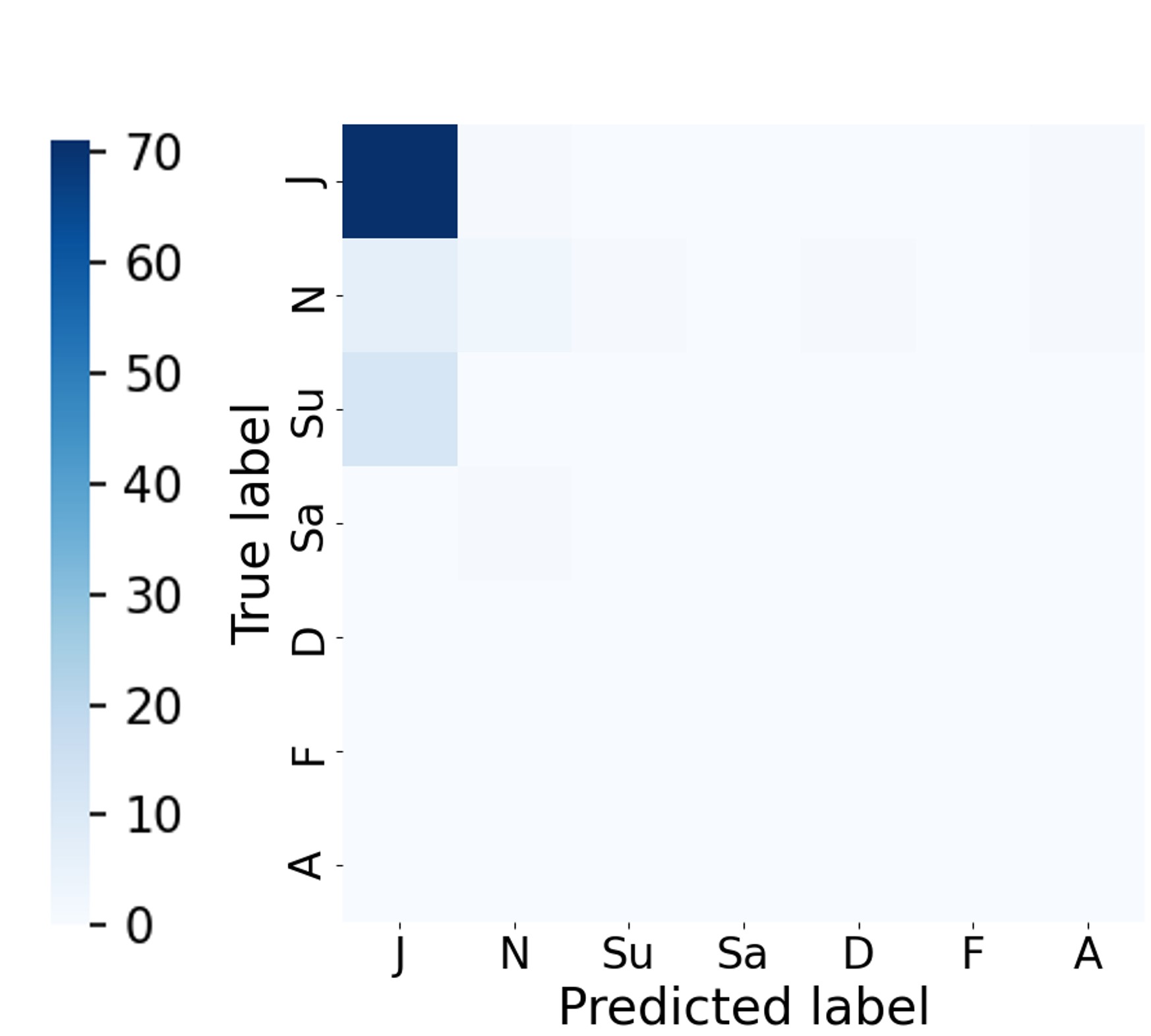}
        \caption{FACET}
        \label{fig:zerocontext}
    \end{subfigure}
    \begin{subfigure}[b]{0.22\textwidth}
        \centering
        \includegraphics[width=\textwidth]{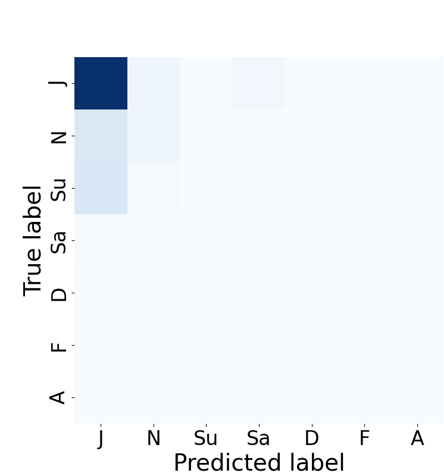}
        \caption{EAC}
        \label{fig:withcont}
    \end{subfigure}
    \begin{subfigure}[b]{0.22\textwidth}
        \centering
        \includegraphics[width=\textwidth]{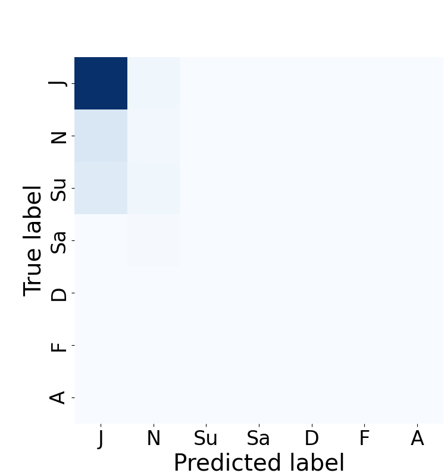}
        \caption{LSTM}
        \label{fig:human_novideo}
    \end{subfigure}
    \caption{Confusion matrix for facial emotion recognition. Human \textit{context-free} is used as a ground truth (\textit{J} stands for Joy, \textit{N} for Neutral, \textit{Su} for Surprise, \textit{Sa} for Sadness, \textit{D} for Disgust, \textit{F} for Fear, and \textit{A} for Anger).}
    \label{fig:confusion_face}
\end{figure*}

\begin{figure*}[h]
    \centering
    \begin{subfigure}[b]{0.22\textwidth}
        \centering
        \includegraphics[width=\textwidth]{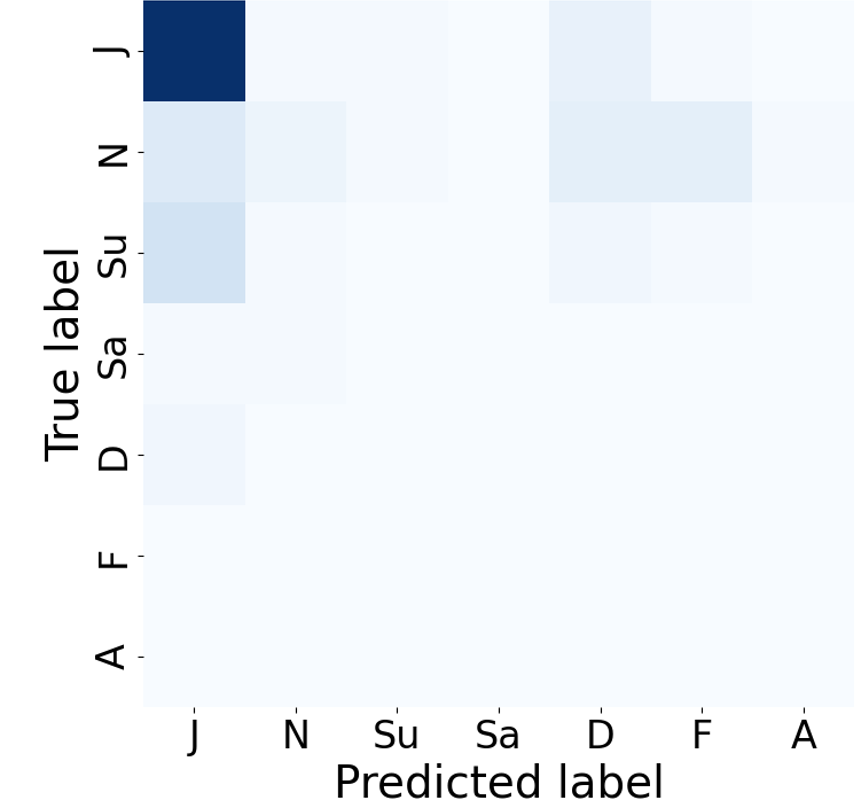}
        \caption{FACET+GPT-3(BCI)}
        \label{fig:zerocontext}
    \end{subfigure}
    \begin{subfigure}[b]{0.22\textwidth}
        \centering
        \includegraphics[width=\textwidth]{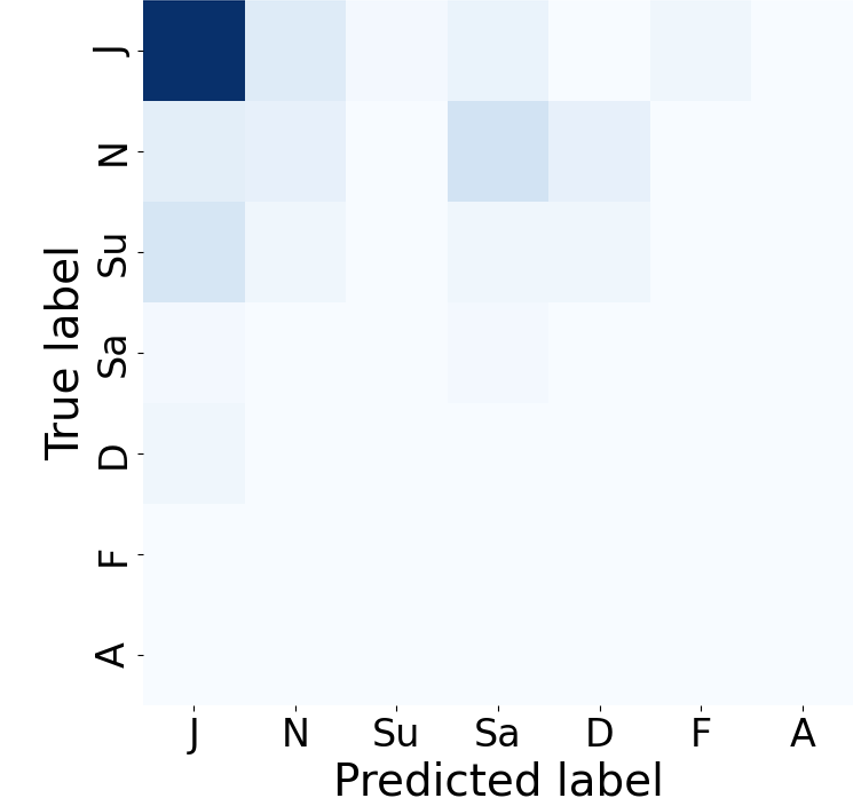}
        \caption{EAC+GPT-3(BCI)}
        \label{fig:withcont}
    \end{subfigure}
    \begin{subfigure}[b]{0.22\textwidth}
        \centering
        \includegraphics[width=\textwidth]{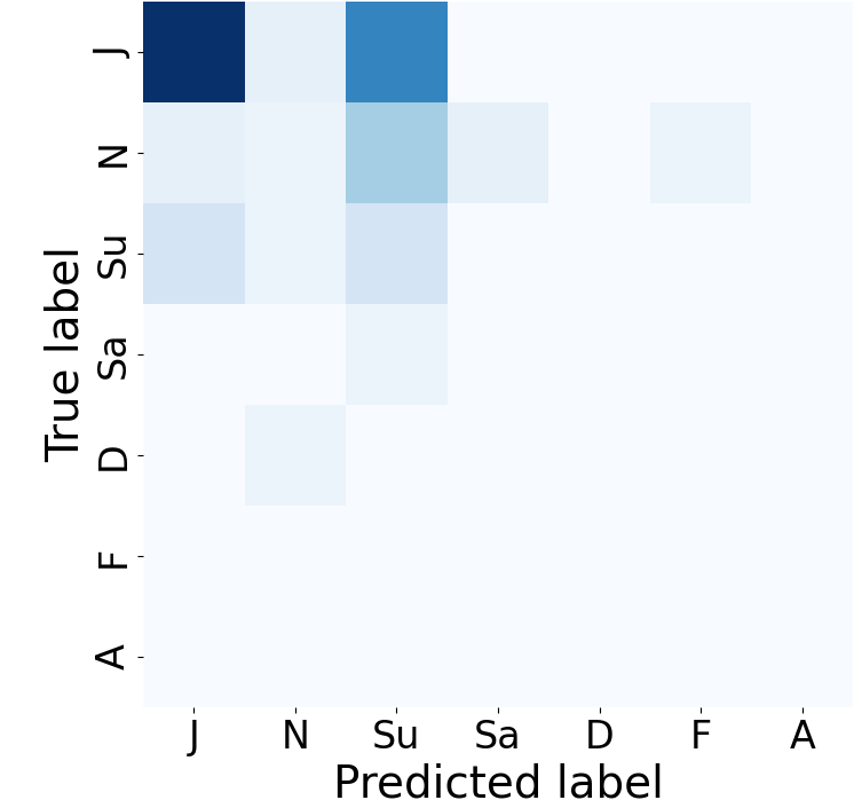}
        \caption{LSTM+GPT-3 (BCI)}
        \label{fig:human_novideo}
    \end{subfigure}
    \\
        \centering
    \begin{subfigure}[b]{0.22\textwidth}
        \centering
        \includegraphics[width=\textwidth]{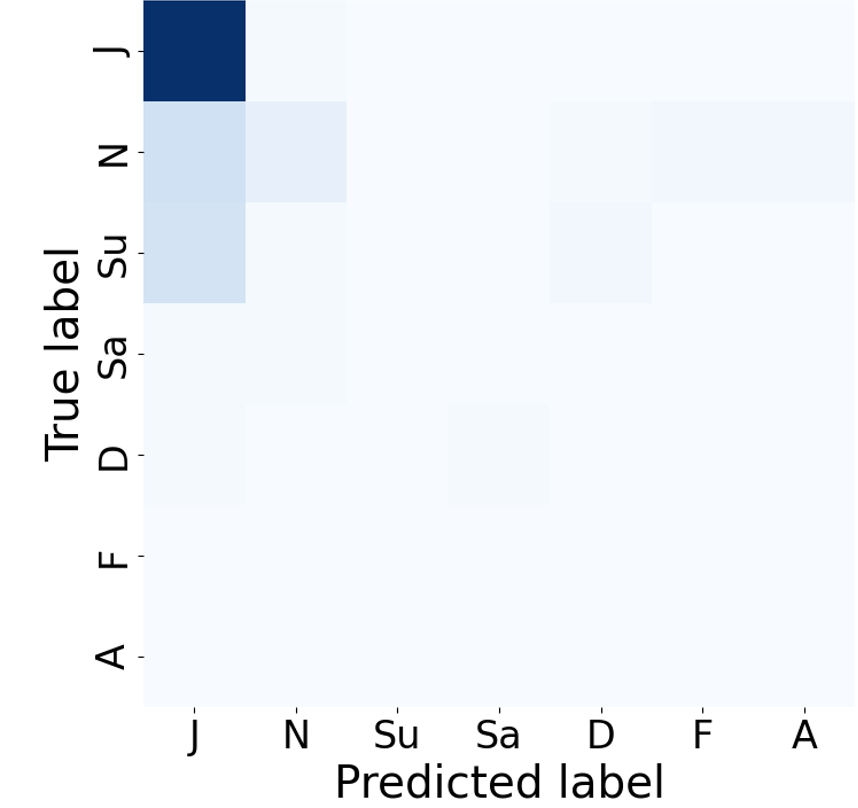}
        \caption{FACET+GPT-4 (BCI)}
        \label{fig:zerocontext}
    \end{subfigure}
    \begin{subfigure}[b]{0.22\textwidth}
        \centering
        \includegraphics[width=\textwidth]{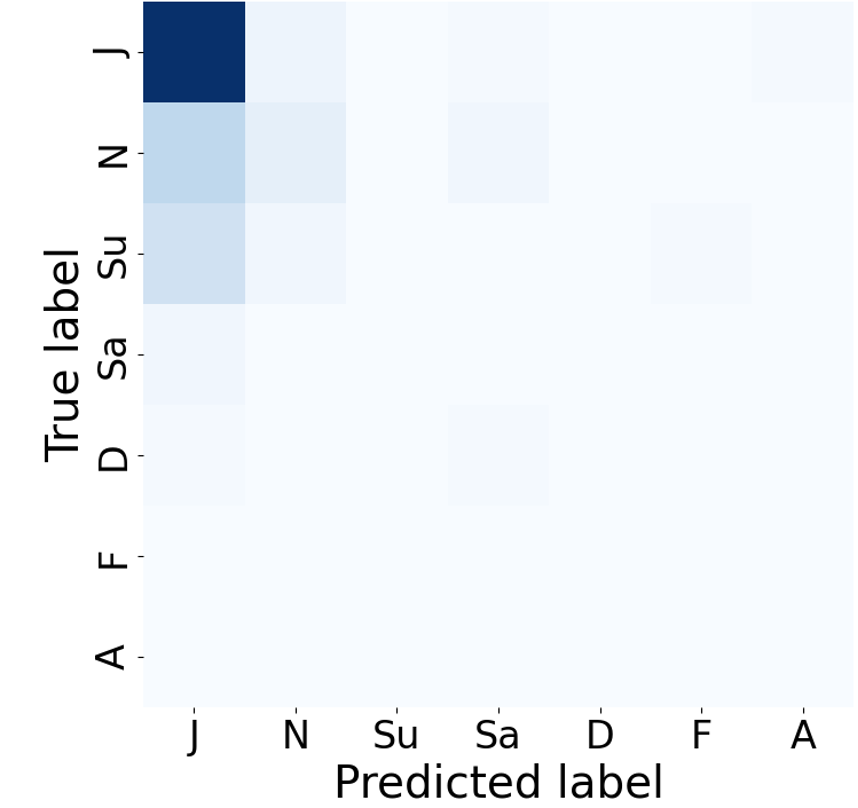}
        \caption{EAC+GPT-4 (BCI)}
        \label{fig:withcont}
    \end{subfigure}
    \begin{subfigure}[b]{0.22\textwidth}
        \centering
        \includegraphics[width=\textwidth]{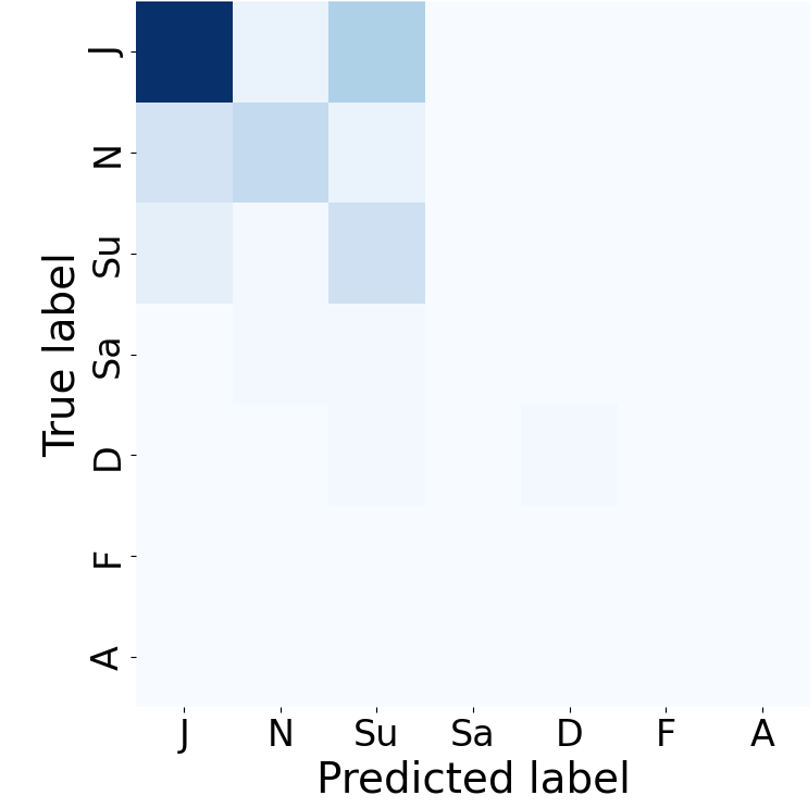}
        \caption{LSTM+GPT-4 (BCI)}
        \label{fig:human_novideo}
    \end{subfigure}
    \\
        \centering
    \begin{subfigure}[b]{0.22\textwidth}
        \centering
        \includegraphics[width=\textwidth]{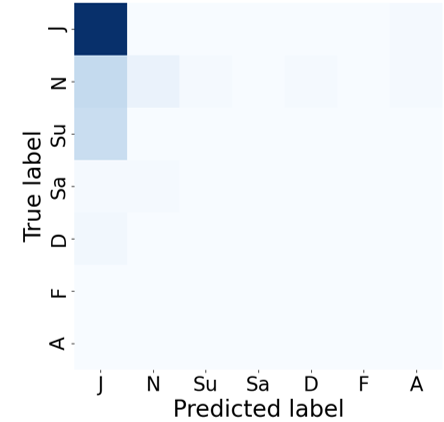}
        \caption{FACET(GPT-4)}
        \label{fig:zerocontext}
    \end{subfigure}
    \begin{subfigure}[b]{0.22\textwidth}
        \centering
        \includegraphics[width=\textwidth]{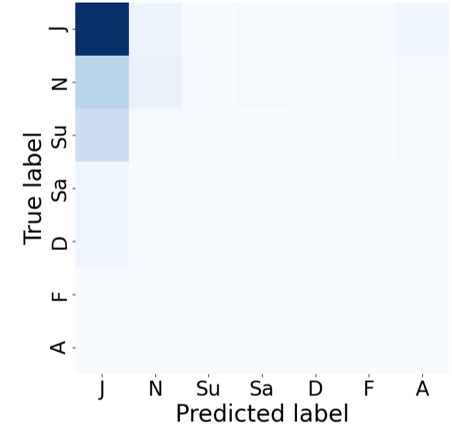}
        \caption{EAC(GPT-4)}
        \label{fig:withcont}
    \end{subfigure}
    \begin{subfigure}[b]{0.22\textwidth}
        \centering
        \includegraphics[width=\textwidth]{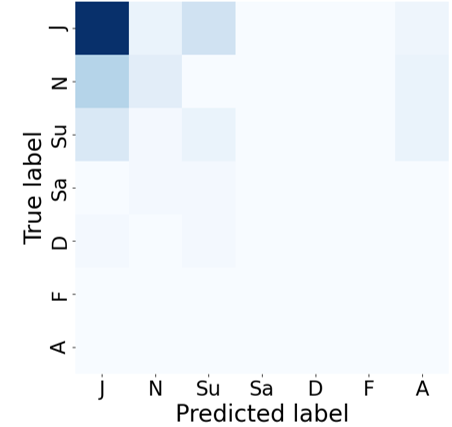}
        \caption{LSTM(GPT-4)}
        \label{fig:human_novideo}
    \end{subfigure}
    \caption{Confusion matrix for facial and contextual emotion recognition. Human \textit{context-based} is used as a ground truth.}
    \label{fig:confusion_all_integration_method}
\end{figure*}

Fig.~\ref{fig:confusion_face} presents the confusion matrix for facial emotion recognition methods using human context-free annotations as the ground truth. All three methods tend to predict emotions as Joy.
Fig.~\ref{fig:confusion_all_integration_method} shows the confusion matrix for facial and contextual emotion recognition. When integrating context, the methods tend to predict a more diverse range of emotions.

\newpage

\subsection{How Integration improves Performance?}

\begin{figure*}[h]
    \centering
    \begin{subfigure}[b]{0.4\textwidth}
        \centering
        \includegraphics[width=\textwidth]{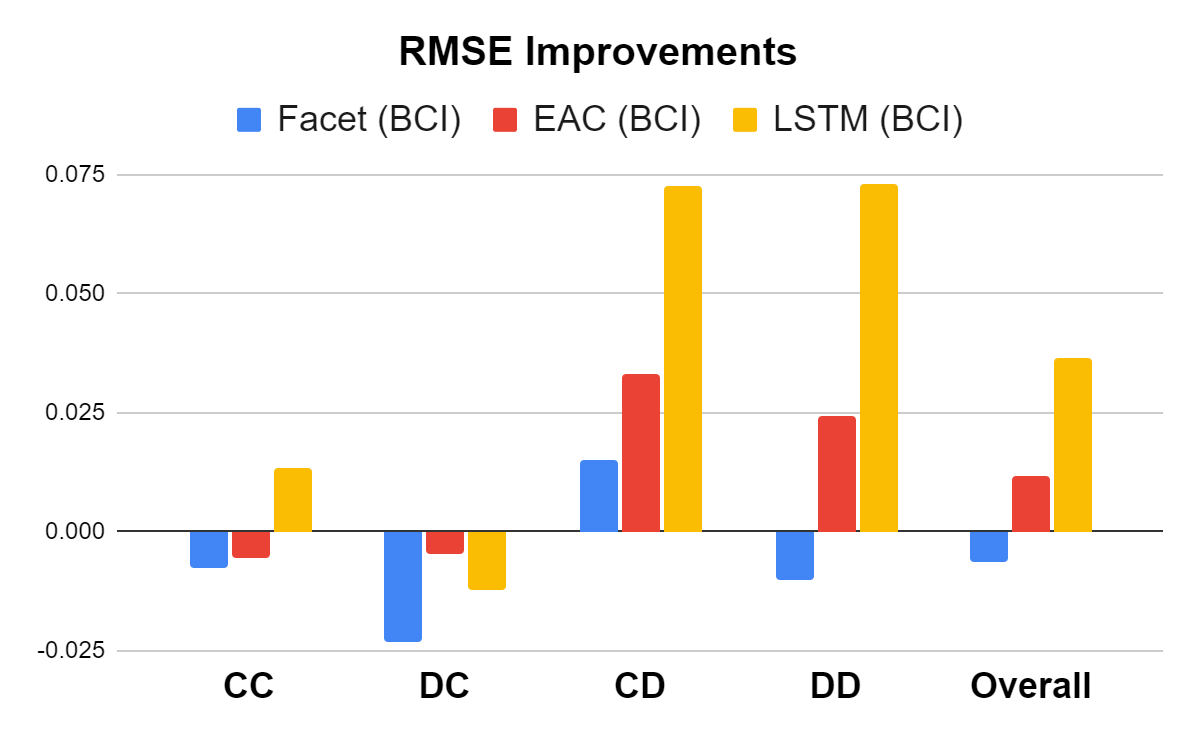}
        \caption{RMSE}
        \label{fig:rmse_improvement}
    \end{subfigure}
    \begin{subfigure}[b]{0.4\textwidth}
        \centering
        \includegraphics[width=\textwidth]{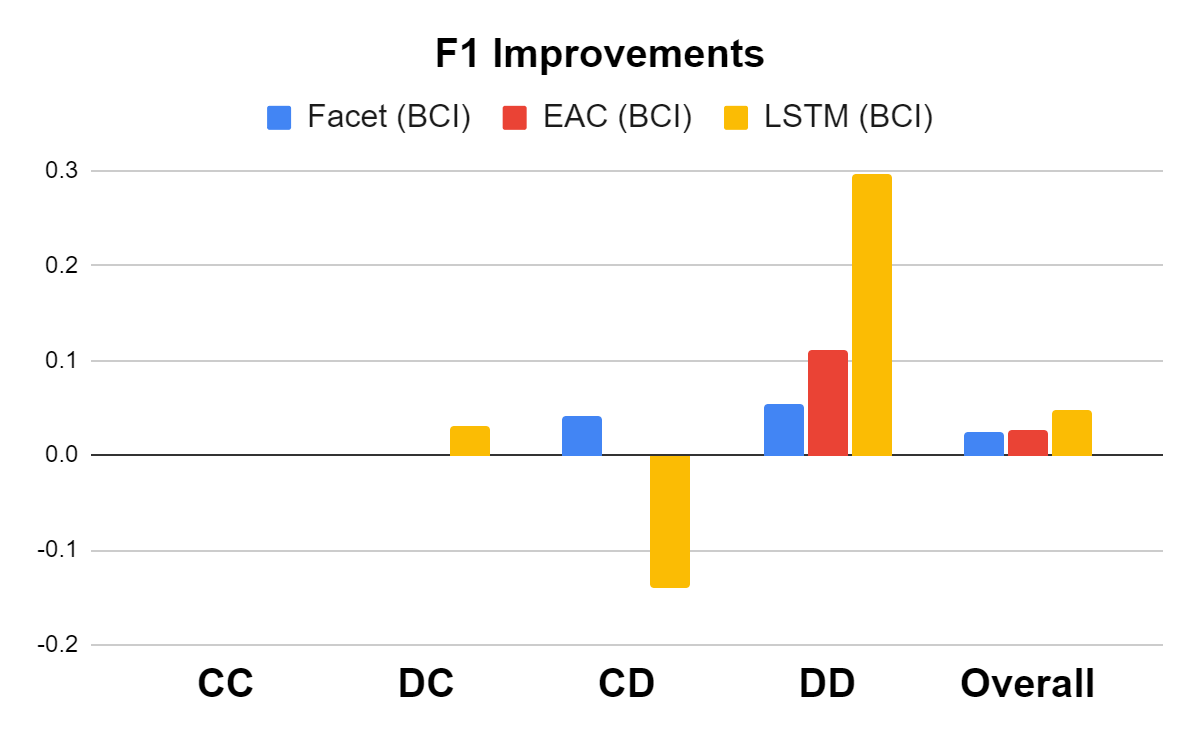}
        \caption{EAC}
        \label{fig:f1_improvement}
    \end{subfigure}
    \caption{Enhancement in model performance (RMSE and KLD) through facial and context integration.}
    \label{fig:Enhancement}
\end{figure*}

We conduct an analysis to assess how BCI enhances recognition performance. Fig.~\ref{fig:Enhancement} illustrates the variation in recognition performance according to different game outcomes. 

\bibliography{reference}
\bibliographystyle{IEEEtran}

\end{document}